\documentclass[prd,preprint,superscriptaddress,nofootinbib,longbibliography,aps]{revtex4-1}
\usepackage[utf8]{inputenc}
\usepackage{comment}
\usepackage{soul}
\usepackage{graphicx}
\usepackage{longtable}
\usepackage{amsmath}
\usepackage{color} 
\usepackage{amssymb}
\usepackage{subfigure}
\usepackage{tabularx}
\usepackage{microtype}
 \usepackage[linktoc=all]{hyperref}
 \hypersetup{
    colorlinks=true,
    linkcolor=BlueViolet,
    citecolor=Blue,
    urlcolor=MidnightBlue,
}

 \usepackage{cleveref}
\usepackage{stackengine}
\usepackage[normalem]{ulem}
\usepackage{slashed}
\usepackage{color}
\usepackage[dvipsnames]{xcolor}

\usepackage{bm}
\usepackage{braket}
\usepackage{slashed}
\usepackage{comment}
\definecolor{myblue}{rgb}{ 0.988, 0.078,0.458}

\graphicspath{{Figures}}

\begin{document}  
\preprint{IFT-UAM/CSIC-25-53}

\title{First astrometric constraints on parity-violation in the gravitational wave background}

\author{Santiago Jaraba}
\email{santiago.jaraba-gomez@astro.unistra.fr}
\affiliation{Observatoire astronomique de Strasbourg, CNRS, Université de Strasbourg, 11 rue de l'Université, 67000 Strasbourg, France}

\author{Sachiko Kuroyanagi}
\email{sachiko.kuroyanagi@csic.es}
\affiliation{Instituto de Física Teórica UAM-CSIC, Universidad Autónoma de Madrid, Cantoblanco 28049 Madrid, Spain}
\affiliation{Department of Physics, Nagoya University, Chikusa, Nagoya 464-8602, Japan}
 
\author{Qiuyue Liang}
\email{qiuyue.liang@ipmu.jp}
\affiliation{Kavli Institute for the Physics and Mathematics of the Universe (WPI), University of Tokyo, Kashiwa 277-8583, Japan}

\author{Meng-Xiang Lin}
\email{mxlin@sas.upenn.edu}
\affiliation{Center for Particle Cosmology, Department of Physics and Astronomy, University of Pennsylvania, Philadelphia, Pennsylvania 19104, USA} 

\author{Mark Trodden} 
\email{trodden@upenn.edu}
\affiliation{Center for Particle Cosmology, Department of Physics and Astronomy, University of Pennsylvania, Philadelphia, Pennsylvania 19104, USA}

\date{\today}

\begin{abstract} 
Astrometry, the precise measurement of stellar positions and velocities, offers a promising approach to probing the low-frequency stochastic gravitational wave background (SGWB). 
Notably, astrometric vector sky maps are sensitive to parity-violating SGWB signals, which cannot be distinguished using pulsar timing array observations in an isotropic SGWB. We present the first astrometric constraints on parity-violating SGWB using quasar catalogs from Gaia DR3 and VLBA data. 
By analyzing the $EB$ correlation in the two-point correlation function of the proper motions of the quasars, we find 2$\sigma$ constraints on the parity-violating SGWB amplitude $h_{70}^2\Omega_{V} = -0.020 \pm 0.025$ from Gaia DR3 and $h_{70}^2\Omega_{V} = -0.004 \pm 0.010$ from VLBA.
These constraints are valid in the frequency range $4.2 \times 10^{-18}\,{\rm Hz} < f < 1.1 \times 10^{-8}\,{\rm Hz}$. 
Although not currently a tight constraint on theoretical models, this first attempt lays the groundwork for future investigations using more precise astrometric data.
\end{abstract} 

\maketitle

\section{Introduction} 
Precision cosmology and astronomy have opened up the exciting possibility of testing fundamental physics through observations of the universe. A particularly interesting prospect is that of searching for violations of basic symmetries of nature. 

A fascinating example is parity violation which, if present in gravitational or other physics in the early universe, might give rise to a range of different phenomena. This possibility has been explored in a number of recent papers that exploit birefringence in the Cosmic Microwave Background (CMB)~\cite{Minami:2020odp,Diego-Palazuelos:2022dsq,Eskilt:2022cff} and measurements of the galaxy four-point correlation function to provide constraints on parity-violating physics~\cite{Hou:2022wfj,Philcox:2022hkh}. In this paper, we focus on the potential signatures of parity violation in gravitational waves (GWs).

Recent reports from pulsar timing array (PTA) collaborations~\cite{NANOGrav:2023gor,EPTA:2023fyk,Reardon:2023gzh,Xu:2023wog,InternationalPulsarTimingArray:2023mzf,Miles:2024seg} indicate a signal consistent with a nanohertz stochastic gravitational wave background (SGWB) at a statistical significance of approximately $3\sigma$. 
A definitive claim of this discovery requires more precise measurements and longer observation times. However, as has been the case with the discovery of other backgrounds in the universe, such as the CMB, the most significant consequence of this finding may be its potential as a unique probe of new physics. Beyond its fundamental importance, applying these results to provide constraints on parity violation may also offer insights into the unresolved question of the origin of any SGWB.

Astrophysically, the SGWB could arise from supermassive black hole binaries emitting nanohertz GWs during their inspiral phase. From a cosmological perspective, GWs generated in the early universe, such as those produced by inflation, alternative cosmological models, or phase transitions occurring before Big Bang Nucleosynthesis (BBN), could also contribute to this background~\cite{NANOGrav:2023hvm,EPTA:2023xxk}.  
Under the assumption of general relativity, the astrophysical background is expected to exhibit only a subdominant level of parity violation, arising from Poisson fluctuations in the number of unresolved sources~\cite{ValbusaDallArmi:2023ydl}. However, parity-violating theories -- such as Chern-Simons gravity, parity-violating extensions of ghost-free scalar-tensor theory, symmetric teleparallel gravity, and Horava-Lifshitz theory -- could lead to modifications in both the generation and propagation of GWs in compact binaries and predict a parity violating astrophysical background~\cite{Alexander:2007kv,Yagi:2013mbt,Callister:2023tws}. For cosmological origins, example models that might produce parity violating signals in the nanohertz frequency band include a pseudoscalar field interacting with a U(1) gauge field and rolling for several e-folds during inflation~\cite{Namba:2015gja,Garcia-Bellido:2023ser}, a parity-violating magnetogenesis model~\cite{Okano:2020uyr}, and a Chern-Simons pseudoscalar active during inflation~\cite{Alexander:2024klf}. Thus, detecting parity violation in the SGWB could provide critical insights into the fundamental nature of gravity and high-energy physics.

One obstacle to this program is that, as pointed out in~\cite{Kato:2015bye,Smith:2016jqs,Liang:2023pbj}, PTA systems are not sensitive to parity information if the SGWB is isotropic. In such cases, the influence of GWs on pulsar measurements is entirely represented by a scalar quantity -- the residuals of photon arrival times -- in the all-sky map. Since a scalar map alone cannot capture parity information, additional data are necessary to tackle this limitation. One potential solution is to relax the assumption of an isotropic background. Anisotropy in the SGWB could provide directional information on the sky map, thereby allowing for the extraction of parity information~\cite{Belgacem:2020nda,Tasinato:2023zcg,Cruz:2024esk}. Another approach involves the concept of a pulsar polarization array~\cite{Liu:2021zlt}. By measuring the polarization vectors of pulsar signals, it is possible to construct a vector map that captures parity information effectively. 

The focus of this paper is on astrometry, which provides independent and complementary constraints on the SGWB beyond those obtained from PTAs. Astrometry employs precise measurements of the positions and motions of celestial objects from modern missions such as Gaia and VLBA, and offers a powerful framework for probing the SGWB in the same frequency range as PTAs~\cite{Pyne:1995iy,Kaiser:1996wk,Gwinn:1996gv,Jaffe:2004it,Book:2010pf}. Notably, it has been proposed that astrometry is also sensitive to a parity-violating background~\cite{Liang:2023pbj}. Assuming an isotropic SGWB with a parity-violating component in the nanohertz band, the idea is to decompose the deflection vectors of the individual proper motions of stellar objects into vector spherical harmonics to detect a non-vanishing $EB$ correlation (i.e. a correlation between the parts of the vector field analogous to the electric and magnetic parts of an electromagnetic field) in the two-point correlation function of stellar position deflections.

In this paper, we present the first analysis of the third Gaia data release (DR3)~\cite{Gaia:2016zol,Gaia:2022,Gaia:2022val} and VLBA~\cite{Truebenbach:2017nhp} data to obtain astrometric constraints on parity-violating SGWB. 
We analyze the same quasar-based datasets as those that have been used to provide the upper limit on the parity-even isotopic SGWB for Gaia DR3~\cite{Jaraba:2023djs} and VLBA~\cite{Darling:2018hmc}. From the spheroidal-toroidal (namely $EB$) correlation of the proper motions in these datasets, we use the quadrupole, as the leading-order term, to set our constraints. Although this technique does not presently yield tight constraints on theoretical models, this analysis should serve as a template for the analysis of future datasets, ultimately aiming to detect parity violation and to explore new fundamental physics in the gravitational sector. We also compare our results to those previously obtained for a parity-even SGWB, and address the potential of future data releases to further refine our constraints.

Before presenting the details of our analysis, we would like to highlight another difference between PTAs and astrometry regarding their respective abilities to place bounds on GWs at lower frequencies. Pulsar timing measurements are most sensitive to GWs at frequencies inversely related to the observation time $T_{\rm obs}$. This is also true for astrometry when using time-series measurements of proper motions, which are expected to be provided in Gaia DR4. 
To claim a detection of GWs in the future, it is essential to observe at least one cycle of GW oscillation using time-series measurements, setting the minimum observable frequency range to be greater than $T_{\rm obs}^{-1}$. However, current DR3 data only provide time-averaged proper motion values. Nevertheless, constraints on the amplitude of GWs can still be established by placing an upper bound for very low frequency GWs ($f< T_{\rm obs}^{-1} \sim 10^{-8}$ Hz), since these induce a quadrupole pattern in proper motions that evolves linearly over time. For frequencies that are lower than the inverse of the minimum distance among the sources --- such as the Hubble distance for quasars --- these GWs do not produce noticeable quadrupole patterns. As a result, astrometric measurements are insensitive to frequencies below $H_0\sim   10^{-18}$Hz.
For frequencies that are within the sensitivity range, such GWs contribute equally to astrometric measurements and the quantity that observations will constrain is the total $\int_{f<T_{\rm obs}^{-1}} df\, \ln (f) ~ \Omega_{\rm GW}(f)$~\cite{Book:2010pf}. 
Thus the resulting bound on the amplitude of GWs within the sensitivity range is frequency-independent.
Note that this argument does not apply to standard PTA observations, since the linear evolution of the timing residuals is typically subtracted when determining the intrinsic spin-down rate of each pulsar. Consequently, a dedicated analysis is required to constrain the lower frequency range using PTA measurements, and sensitivity drops with a dependency $\Omega_{\rm GW}\propto f^{-2}$ as we move to these lower frequencies~\cite{Yonemaru:2017esn,Kumamoto:2019uoy,DeRocco:2022irl,DeRocco:2023qae}.

The outline of this paper is as follows. In Sec.~\ref{sec:formula}, we introduce the observable describing parity violation in an astrometric system. We compute the proper motion $EB$ correlation using the quadrupole approximation and derive the relation between the signal and the amplitude. In Sec.~\ref{sec:data}, we discuss the methodology and conduct the data analysis, presenting the $1\sigma$ and $2\sigma$ bounds on parity violating signals in the Gaia DR3 and VLBA data sets. We then discuss our results and conclude in Sec.~\ref{sec:conlusion}. 

\section{Parity violating effects in astrometry} 
\label{sec:formula}
In this section, following~\cite{Liang:2023pbj,Book:2010pf}, we begin by reviewing the formalism through which parity-violating effects can be measured using astrometry. We then show how to express the resulting observable in terms of Stokes parameters, and connect this to quantities conventionally used in GW physics. 

We begin by writing a plane GW as 
\begin{equation}
    h_{i j}(t, \vec{x})=\sum_{S=+, \times} \int \frac{\mathrm{d}^3 \vec{k}}{(2 \pi)^3} h_S(t, \vec{k}) e_{i j}^S(\hat{k}) e^{i \vec{k} \cdot \vec{x}} = \sum_{S=+, \times} 
    \int d^2\hat{k} 
    \int_0^\infty \frac{d k}{2 \pi} \frac{k^2}{(2 \pi)^2} 
    h_S(t, \vec{k}) e_{i j}^S(\hat{k}) e^{i \vec{k} \cdot \vec{x}} \ ,
\end{equation}
where $S$ denotes the $+,\times$ polarization mode, $k = |\vec k| = 2\pi f$ is the absolute value of the momentum, $\hat k = \vec k/k$ is the direction of the propagation,
and the polarization tensors satisfy $e_{i j}^S(\hat{k})e^{S' i j}(\hat{k}) = 2 \delta^{SS'}$.
The plane wave assumption guarantees that
\begin{eqnarray}
    h_S(t, \vec{k}) = e^{-i k t} h_S(k,\hat k)\ .
\end{eqnarray}
 
The SGWB can be described by the two-point correlation functions in Fourier space~\cite{Liang:2023pbj,Kato:2015bye}
\begin{eqnarray}
     \left\langle h_S^*(t, \vec{k}) h_{S^{\prime}}(t, \vec{k}^{\prime})\right\rangle=(2 \pi)^3 \delta^{(3)}(\vec{k}-\vec{k}^{\prime}) P_{S S^{\prime}}(k) =  \frac{\delta(f-f')}{f^2}\delta^{(2)}(\hat k,\hat k') P_{SS'}(k)\ ,
\end{eqnarray}
\begin{equation}
\label{eq,stokes}
 \begin{aligned}
& { P_{S S^{\prime}}(k) =\left(\begin{array}{ll}
P_{++}(k) & P_{+\times}(k) \\
P_{\times+}(k) & P_{\times \times}(k)
\end{array}\right)} =\frac{1}{2}\left(\begin{array}{cc}
I+Q & U-i V \\
U+i V & I-Q
\end{array}\right)_{SS^{\prime}}
\end{aligned}  \ ,
\end{equation}
where $[P_{SS'}(k)] = [k^{-3}]$ is the power spectrum\footnote{For an isotropic background, we can relate $P_{SS'}(k)$ to the spectral density quantity $S_h(f)$ in~\cite{Maggiore:1999vm} by $P_{++}(k)=P_{\times\times}(k) = \frac{S_h(f)}{4\pi f^2}$. }. Here, $I$, $U$, $V$, and $Q$ are the Stokes parameters~\cite{Kato:2015bye}. The parameter $I$ represents the intensity of the GW and corresponds to the parity-even part of the signal. 
The $U$ and $Q$ parameters characterize the linear polarization state and vanish for an isotropic background. The parameter $V$ quantifies the difference between the left-moving and right-moving modes 
in the circular polarization basis that is defined by $e^{R}_{ij}=(e^{+}_{ij}+ i e^{\times}_{ij})/\sqrt{2}$ and $e^{L}_{ij}=(e^{+}_{ij}- i e^{\times}_{ij})/\sqrt{2}$,
thereby characterizing the asymmetry in circular polarization amplitudes.

The energy density and the fractional energy density of GWs are then given 
by~\cite{Maggiore:1999vm}~\footnote{Note that there is a factor of two in comparison with the results in~\cite{Maggiore:1999vm} that arises from a summation over polarization states.}
\begin{eqnarray}
   \rho_{\mathrm{gw}}&=& \frac{1}{32 \pi G}\left\langle\dot{h}_{i j}(t, \vec{x}) \dot{h}^{i j}(t, \vec{x})\right\rangle = \frac{1}{32 \pi G} \sum_{S=S'} \int_0^\infty df ~ 32 \pi^3  f^4 P_{SS'}
   \ ,\label{eq:rhoGW}\\
   \Omega_{\mathrm{gw}}(f)&=& \frac{1}{\rho_c} \frac{d \rho_{\mathrm{gw}}}{d \log f}  =\frac{8\pi^3}{3H_0^2} \sum_{S=S'}f^5P_{SS'}
   = \frac{8\pi^3}{3H_0^2}\sum_{S=S'}\frac{f^2 \mathcal{A}_{SS'}} {\Delta \ln f}   \ ,
   \label{eq:OmegaGW}
\end{eqnarray}
where $\rho_c = 3H_0^2/8\pi G$ is the critical density of the universe. Here we have used $k = 2\pi f$ to obtain the equivalent momentum-space expression, and $\mathcal{A}_{SS'}$ is the dimensionless amplitude defined as
\begin{eqnarray}
\label{eq:amplitude}
     \mathcal{A}_{SS'}=\int \mathrm{d} k \frac{k^2}{(2 \pi)^3} P_{S S^{\prime}}(k)  \approx \frac{k^3\Delta \ln k}{(2 \pi)^3} P_{S S^{\prime}}(k) \ .
\end{eqnarray}  
As discussed later, $\mathcal{A}_{SS'}$ can be related to the angular deflection angle of the stellar object, and all frequencies (or momenta) with $f < T_{\rm obs}^{-1}$ contribute to the induced proper motion. However, it is more convenient to provide observational constraints for each logarithmic frequency bin. Therefore, in the second step of Eq.~\eqref{eq:amplitude}, we approximate the GW contribution as coming from the interval between $\ln k$ and $\ln k + \Delta \ln k$.

It is then convenient to define the matrix quantity
\begin{eqnarray}
    \label{eq:Omega_crossplus}
  \Omega_{ SS' }(f)   \equiv \frac{8\pi^3}{3H_0^2}  \frac{f^2 \mathcal{A}_{SS'}} {\Delta \ln f} \ ,
\end{eqnarray}
for which the diagonal components $\Omega_{ + + }(f)$ and $\Omega_{ \times \times}(f)$ correspond to the fractional energy densities in GWs with $+$ and $\times$ polarizations, respectively, and contribute in Eq.~\eqref{eq:OmegaGW}. Note that the off-diagonal components are expressed as combinations of the $U$ and $V$ components and do not directly correspond to their energy densities.
Equivalently, we can define
\begin{equation}
    \label{eq:omega_u}
    \Omega_U = 2{\rm Re}(\Omega_{\times +}),\quad\Omega_V = 2{\rm  Im }(\Omega_{\times +})\ ,
\end{equation}
to analogously describe the fractional energy densities of $U$ and $V$.

The angular deflection angle that is induced by the GW is
~\cite{Liang:2023pbj}
\begin{equation}
    \label{eq:delta_n}
    \delta n^I(t, \hat{n})=e^I{ }_\mu \delta n^\mu(t, \hat{n})=\mathcal{R}^{I J K}(\hat{n}, \hat{k}) h_{J K}(\vec{k}) e^{-i k t} \ ,
\end{equation}
where 
\begin{eqnarray}
    \mathcal{R}^{I J K}(\hat{n}, \hat{k})=\frac{\hat{n}^I+\hat{k}^I}{2(1+\hat{k} \cdot \hat{n})} \hat{n}^J \hat{n}^K-\frac{1}{2} \delta^{I J} \hat{n}^K \ ,
\end{eqnarray}
with $\hat{n}$ being the line of sight unit vector from the observer to the stellar object.
Then, the two-point correlation function can be expressed as
\begin{eqnarray}
\left\langle\delta n^I(t, \hat{n}) \delta n^J\left(t, \hat{n}^{\prime}\right)\right\rangle &=& 
\sum_{S, S^{\prime}}  \mathcal{A}_{SS'} \int d^2 \hat k  \mathcal{R}^{I K L}(\hat{n}, \hat{k}) \mathcal{R}^{J M N}(\hat{n}^{\prime}, -\hat{k}) e_{K L}^S(\hat{k}) e_{M N}^{S^{\prime}}(-\hat{k}) \ ,
\end{eqnarray}
where $ \mathcal{A}_{SS'}$ is the dimensionless amplitude defined in Eq.~\eqref{eq:amplitude}, and the remaining part denotes the angular correlation between quasar pairs in the astrometry system. The integration over $d^2 \hat k $ accounts for the contributions from GWs arriving from all directions. 

The parity violation term $\mathcal{A}_{\times +} =\mathcal{A}_{+\times}^*$ contributes to the two-point correlation matrix among different stellar objects. It is convenient to describe this correlation by decomposing the angular deflection angle in Eq.~\eqref{eq:delta_n} in vector spherical harmonics, as
\begin{equation}
    \label{eq:delta_n_decomp}
    \delta \vec{n}(t, \hat{n})=\sum_{\ell m}\left[\delta n_{E \ell m}(t) \vec{Y}_{\ell m}^E(\hat{n})+\delta n_{B \ell m}(t) \vec{Y}_{\ell m}^B(\hat{n})\right] \,,
\end{equation}
where $E,B$ stand for the spheroidal (or electric) and toroidal (or magnetic) modes, respectively. 
The $EB$ correlation is proportional to $\mathcal{A}_{\times+}$ for even $\ell$ and to $\mathcal{A}_{+\times}$ for odd $\ell$, which vanish in a parity conserving theory. We only use the quadrupole contribution, as it is the dominant term~\cite{Liang:2023pbj}
\begin{eqnarray}
    \label{eq:delta_n_A}
    \sum_{m=-2}^2\left\langle\delta n_{E 2 m} \delta n_{B 2 m}^*\right\rangle = \frac{20 \pi^2}{9} \mathcal{A}_{\times+} \,.
\end{eqnarray}
Analyzing astronomical data by directly applying this equation is challenging, since computing the angular deflection angle of a given astronomical object involves defining its unperturbed position. However, working with the proper motions provided by astrometric surveys, as it is done in~\cite{Jaraba:2023djs} and~\cite{Darling:2018hmc}, removes the need for such definition.  Moreover, since $\delta\vec{n}$ is a real-valued vector field, we can express the time derivative of Eq.~\eqref{eq:delta_n_decomp} as
\begin{align}
    \label{eq:pm_sph}
    \delta\dot{\vec{n}}(\hat{n})=&\sum_{l=1}^\infty\Biggl(\delta \dot{n}_{El0}\vec{Y}^E_{l0}(\hat{n})+\delta \dot{n}_{Bl0}\vec{Y}^B_{l0} (\hat{n})\\
    \nonumber
    &+\left.\sum_{m=1}^l\left[\delta \dot{n}_{Elm}^{\rm Re}\vec{Y}^{E,\rm Re}_{lm}(\hat{n})+\delta \dot{n}_{Blm}^{\rm Re}\vec{Y}^{B,\rm Re}_{lm}(\hat{n})-\delta \dot{n}_{Elm}^{\rm Im}\vec{Y}^{E,\rm Im}_{lm}(\hat{n})-\delta \dot{n}_{Blm}^{\rm Im}\vec{Y}^{B,\rm Im}_{lm}(\hat{n})\right]\right)\,,
\end{align}
where the superscripts Re and Im stand, respectively, for the real and imaginary parts of the coefficients or spherical harmonics basis.

In a similar way to the analyses in~\cite{Jaraba:2023djs,Darling:2018hmc}, we define the quadrupole power for the mixed modes as
\begin{eqnarray}
    \label{eq:Ppluscross}
  \tilde P_2^{\times+}\equiv \sum_{m=-2}^2\left\langle\delta \dot{n}_{E 2 m} \delta \dot{n}_{B 2 m}^*\right\rangle=\frac{20 \pi^2}{9} k^2 \mathcal{A}_{\times+} \ ,
\end{eqnarray}
where the last step follows from Eq.~\eqref{eq:delta_n_A}. Note that here, $\tilde P_2^{\times+} $ denotes the mass dimension 2 quadrupole power, which is distinct from the power spectrum $P_{SS'}$. From Eq.~\eqref{eq,stokes}, we know that the imaginary part of $\tilde P_2^{\times+}$ is related to the Stokes parameter $V$, while its real part is related to the Stokes parameter $U$, which encodes additional information about linear polarizations and whihc vanishes under the assumption of an isotropic SGWB.

Substituting this result into Eq.~\eqref{eq:delta_n_A} and Eq.~\eqref{eq:Omega_crossplus}, we obtain
\begin{eqnarray}
    \label{eq:Omega_pluscross_final}
    \Omega_{\times+ }(f) 
    = \frac{3}{10\pi} \frac{\tilde P_2^{\times+}}{H_0^2} = 0.000438 \frac{\tilde P_2^{\times+}}{(1~\mu \mathrm{as} / \mathrm{yr})^2} h_{70}^{-2} \ ,
\end{eqnarray}
where we have assumed that the signal is the dominant feature over an order of magnitude in frequency ($\Delta\ln f = 1$), and have fixed the value of the Hubble constant to be $H_0\approx 14.67 h_{70}~\mu\mathrm{as} / \mathrm{yr} $. Also note that, while $  \Omega_{\times+ }(f)$ is a complex number in general, under the assumption of an isotropic SGWB, it is purely imaginary.

\section{data analysis}
\label{sec:data}

The formalism of the previous section allows us to directly apply the posterior probability distributions obtained in~\cite{Jaraba:2023djs} to yield novel constraints.  We refer the reader to~\cite{Jaraba:2023djs} for a detailed description of the methodology and dataset construction. In this section, we summarize the key aspects of the method and datasets used and present our results.

\subsection{Methodology}

In order to set observational constraints, datasets with the positions and proper motions of a set of quasars (QSOs) are used. Then, a generic proper motion field (Eq.~\eqref{eq:pm_sph}) up to the quadrupole moment ($\ell=1,2$) is fitted to this data. 
When no signal is present and we are essentially fitting random noise, higher-order modes do not add useful information.
Despite the fact that only the quadrupole coefficients are needed to derive GW constraints, it is nevertheless important to account for the dipole component, which does not carry any GW information but can bias the derived quadrupole coefficients if it is not treated carefully~\cite{Jaraba:2023djs,Darling:2018hmc}.

The fits are carried out via a Bayesian approach, using the Markov chain Monte Carlo (MCMC) algorithm implemented in the Python package \textit{emcee}~\cite{Foreman-Mackey:2012any}. Generous uniform priors of $[-100, 100]~\mathrm{\mu as}$ were used for each of the 16 coefficients, where 6 coefficients correspond to the dipole (3 for each mode E, B) and 10 to the quadrupole. In addition, a specific statistical treatment to reduce the weight of outlier sources~\citep{Sivia:2006} was used, which, in practice, amounts to substituting the standard, quadratic log-likelihood by
\begin{equation}
    \ln\mathcal{L}={\rm const.}+\sum_{i=1}^N \ln\left(\frac{1-e^{-R_i^2/2}}{R_i^2}\right)\ ,\quad R_i = \frac{D_i - {\rm Model}}{\sigma_i}\ ,
\end{equation}
with $D_i$ being a data point of uncertainty $\sigma_i$. Finally, the convergence of the algorithm was secured by running for a number of iterations greater than 100 times the maximum autocorrelation time. 

The statistical significance of the parameter estimation was quantified through both the Bayes factor and the Z score, which quantifies how many standard deviations separate the data from the pure noise hypothesis (see~\cite{Mignard:2012xm,Jaraba:2023djs} for a detailed definition and statistical properties). 
Since the algorithm does not find any relevant signal in the data and no detection can be claimed, we instead provide constraints at the 95\% credible level. We perform a similar analysis to that in~\cite{Jaraba:2023djs} here, with the particularity that constraints on $\Omega_{U}$ and $\Omega_{V}$ are presented instead of on $\Omega_{\rm gw}=\Omega_{++}+\Omega_{\times\times}$.

\subsection{Datasets}

The datasets considered in~\cite{Jaraba:2023djs} consist of quasar positions and proper motions from either Gaia DR3 or VLBA, the latter being originally from~\cite{Darling:2018hmc,Truebenbach:2017nhp}.

\subsubsection{Gaia DR3 dataset}

In~\cite{Jaraba:2023djs}, four datasets based on Gaia DR3 are produced and analyzed. The reason to consider four datasets was the lack of an official quasar catalog for Gaia DR3 at the time. The Quaia catalog~\cite{Storey-Fisher:2023gca}, released subsequently, has been used for astrometric constraints in~\cite{Darling:2024myz}. Using diverse information from the source classification of extragalactic sources provided by the Gaia collaboration~\cite{Gaia:2022vcs}, the authors in~\cite{Jaraba:2023djs} considered different criteria to generate four selections of sources with high probability of being quasars, aiming to minimize contamination from non-QSO sources. Comparing the results across these datasets demonstrated that the derived constraints were robust and not significantly influenced by the specific source selection.

\begin{figure}[htbp]
    \centering
    \includegraphics[width=0.5\linewidth]{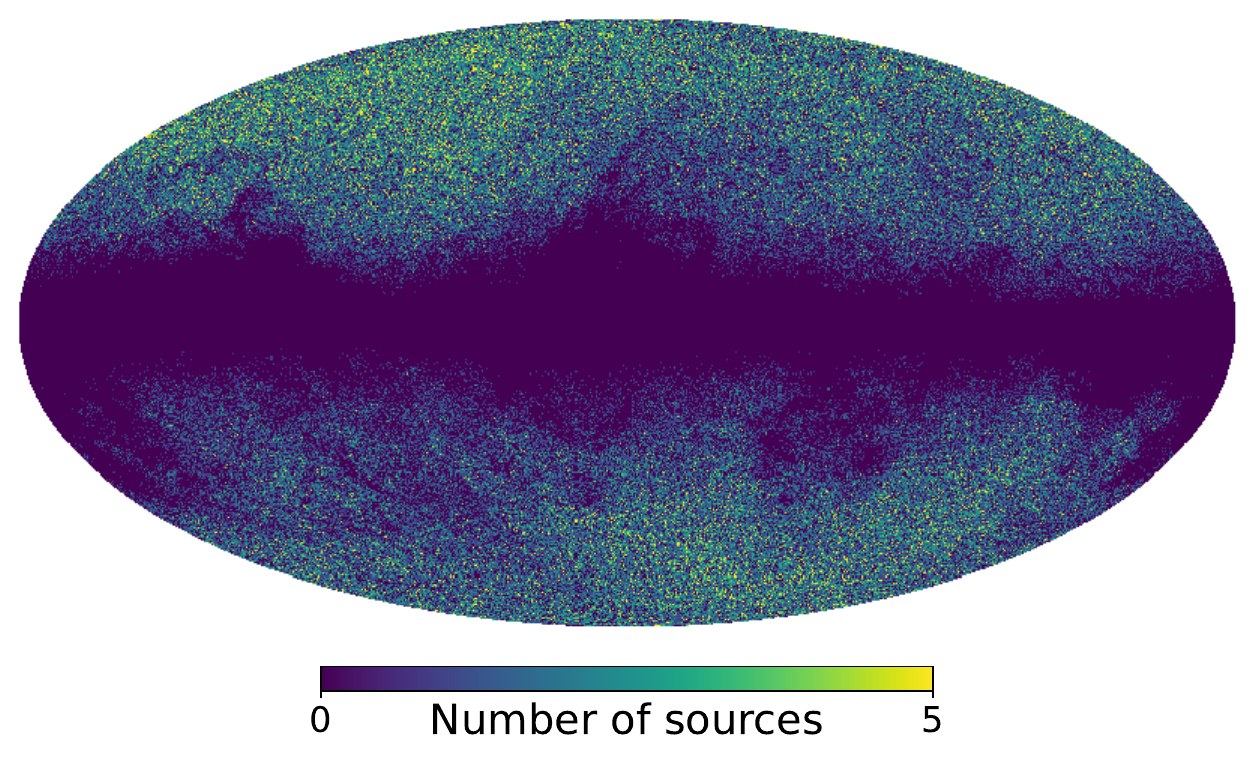}\includegraphics[width=0.5\linewidth]{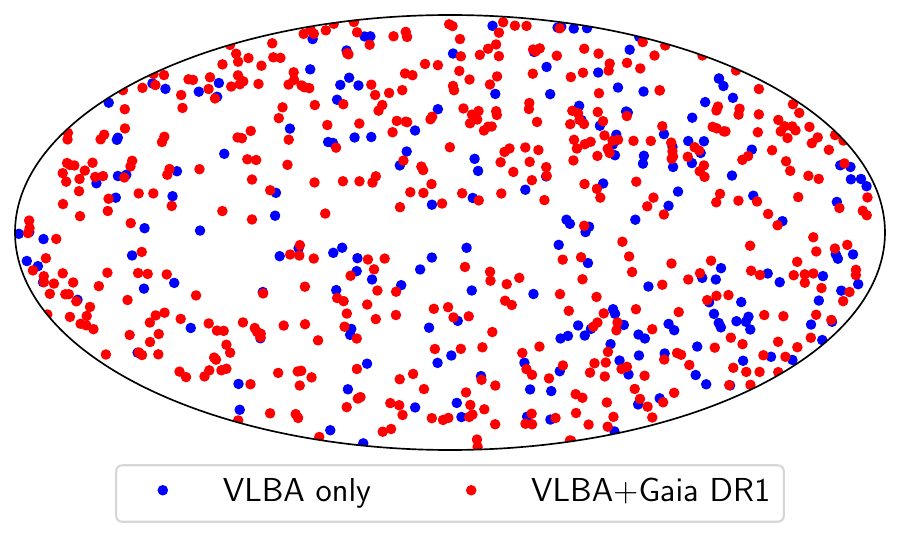}
    \caption{Left: skymap with the number of sources per pixel of the Gaia DR3 dataset, using a HEALPix scheme of level 8 (786,432 pixels, with each of them covering an area of around 189 ${\rm arcmin}^2$. Right: skymap with the positions of all the VLBA (blue and red) and VLBA+Gaia DR1 (red) datasets.}
    \label{fig:sources}
\end{figure}

In this paper, we only use the cleanest dataset among these four (the dataset called ``intersection''), which is obtained by selecting the common sources from the ``pure'' and ``astrometric'' selections provided by Gaia~\cite{Gaia:2022vcs}, and by setting a threshold on the parameter \textit{classprob\_dsc\_combmod\_quasar}, which represents the probability of the source of being a quasar. Further details on how to construct these datasets are provided in~\cite{Jaraba:2023djs}, while the exact queries to download them from the Gaia Archive\footnote{\url{https://gea.esac.esa.int/archive/}} are given in~\cite{JarabaGomez:2024jni}. 
A skymap showing the distribution of these sources is provided in the left panel of Fig.~\ref{fig:sources}, which has been plotted using HEALPix (Hierarchical Equal Area isoLatitude Pixelization)\footnote{\url{https://healpix.sourceforge.io}}.

\subsubsection{VLBA-based datasets}

Despite the more recent launch of Gaia, the Very Long Baseline Array (VLBA) currently offers better constraining power than Gaia DR3 due to its greater observing time (22.2 years for VLBA vs 2.84 years for Gaia DR3). In~\cite{Darling:2018hmc}, two VLBA-based datasets are considered: the proper motions of 711 quasars obtained from VLBA, and updated proper motions for 508 of these sources together with the information from Gaia DR1. In~\cite{Jaraba:2023djs}, these constraints were revisited using the same datasets, showing that the original constraints in~\cite{Darling:2018hmc} were too optimistic. In this work, we consider both of these datasets. The distribution of these sources in the sky is shown in the right panel of Fig.~\ref{fig:sources}.

\subsection{Parity violation constraints}

In this paper, we derive constraints on $\Omega_{U}$ and $\Omega_{V}$ from the posterior probability distributions for the multipole coefficients. Note that $\Omega_{U}$ should vanish for an isotropic SGWB. However, in the presence of anisotropies, a non-zero $\Omega_{U}$ corresponds to the real part of the $EB$ correlation. Therefore, in addition to $\Omega_{V}$, we also provide constraints on $\Omega_{U}$.

To obtain constraints on $\Omega_{U}$ and $\Omega_{V}$, we need to extract the $EB$ correlation, whose posterior distributions have already been obtained in the analysis of~\cite{Jaraba:2023djs}. Therefore, no additional parameter estimation runs are required. The corner plots corresponding to the datasets considered in this work are provided in Appendix~\ref{app:corner}.

In particular, the definitions in Eq.~\eqref{eq:omega_u}, together with Eq.~\eqref{eq:Omega_pluscross_final}, imply
\begin{equation}
h_{70}^2\Omega_{U}=\frac{0.000876}{(1~\mu \mathrm{as} / \mathrm{yr})^2}\left\{\delta\dot{n}_{E20}\delta\dot{n}_{B20}+2\sum_{m=1}^2\left(\delta\dot{n}_{E2m}^{\rm Re}\delta\dot{n}_{B2m}^{\rm Re}+\delta\dot{n}_{E2m}^{\rm Im}\delta\dot{n}_{B2m}^{\rm Im}\right)\right\}\ ,
    \label{eq:OmegaQ}
\end{equation}
\begin{equation}
h_{70}^2\Omega_{V}=\frac{0.000876}{(1~\mu \mathrm{as} / \mathrm{yr})^2}\left\{2\sum_{m=1}^2\left(\delta\dot{n}_{E2m}^{\rm Im}\delta\dot{n}_{B2m}^{\rm Re}-\delta\dot{n}_{E2m}^{\rm Re}\delta\dot{n}_{B2m}^{\rm Im}\right)\right\}\ ,
    \label{eq:OmegaV}
\end{equation}
which directly relate the posteriors shown in Appendix~\ref{app:corner} to the quantities that we aim to constrain.

\begin{figure}[htbp]
    \centering
    \includegraphics[width=0.5\linewidth]{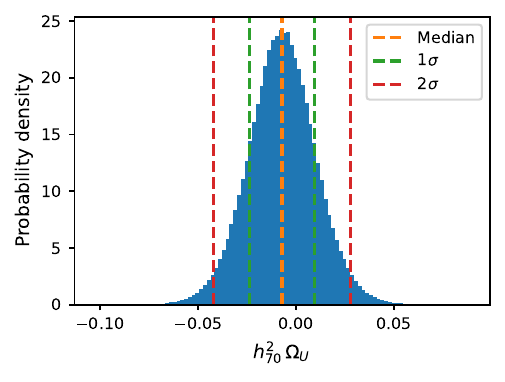}\includegraphics[width=0.5\linewidth]{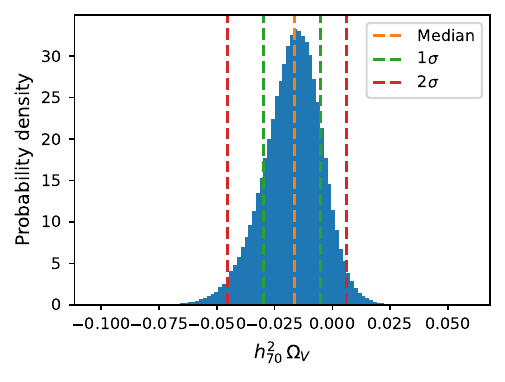}
    \caption{Probability density functions for $h_{70}^2\Omega_{U}$ (left) and $h_{70}^2\Omega_{V}$ (right) computed from the Gaia DR3 dataset. The vertical dashed lines represent the median (orange), the 1$\sigma$ credible interval (green), and the 2$\sigma$ credible interval (red), respectively.
    }
\label{fig:intersection}
\end{figure}

\begin{figure}[htbp]
    \centering
    \includegraphics[width=0.5\linewidth]{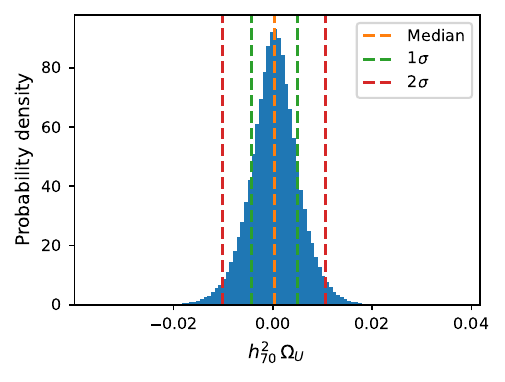}\includegraphics[width=0.5\linewidth]{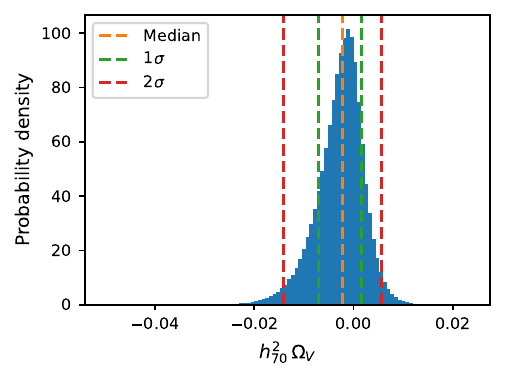}
    \caption{Probability density distributions for $h_{70}^2\Omega_{U}$ (left) and $h_{70}^2\Omega_{V}$ (right) computed from the VLBA dataset. The vertical dashed lines represent the median (orange), the 1$\sigma$ credible interval (green), and the 2$\sigma$ credible interval (red), respectively.
    }
    \label{fig:VLBA}
\end{figure}

\begin{figure}[htbp]
    \centering
    \includegraphics[width=0.5\linewidth]{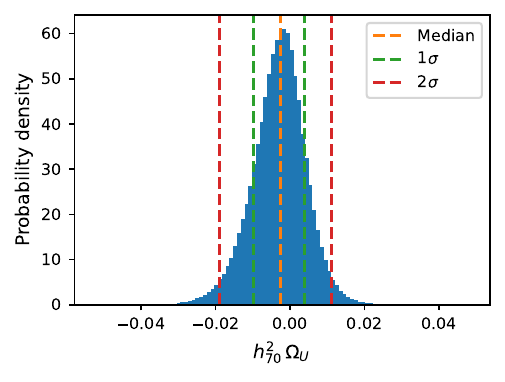}\includegraphics[width=0.5\linewidth]{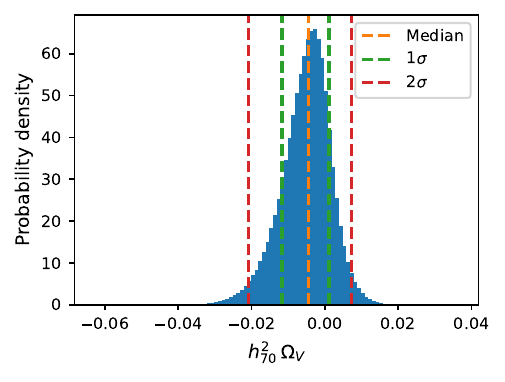}
    \caption{Probability density distributions for $h_{70}^2\Omega_{U}$ (left) and $h_{70}^2\Omega_{V}$ (right) computed from the VLBA+Gaia DR1 dataset. The vertical dashed lines represent the median (orange), the 1$\sigma$ credible interval (green), and the 2$\sigma$ credible interval (red), respectively.
    }
    \label{fig:VLBA_Gaia}
\end{figure}

The probability distributions for these quantities for the Gaia DR3, VLBA and the VLBA+Gaia DR1 dataset are shown in Figs.~\ref{fig:intersection},~\ref{fig:VLBA} and~\ref{fig:VLBA_Gaia}, respectively. Proper motion correlations are translated to $h_{70}^2\Omega_{U}$ and $h_{70}^2\Omega_{V}$ using Eqs.~\eqref{eq:OmegaQ} and \eqref{eq:OmegaV}. 
As expected, the values are consistent with zero since no SGWB detection was found from these data. 
Then the variance of each distribution provides a constraint on the parity-violating SGWB amplitude. The 2$\sigma$ constraints obtained from these distributions are listed in Table~\ref{tab:results}. 
The tightest constraints come from VLBA, which yields $h_{70}^2\Omega_V = -0.004\pm 0.010$ at 2$\sigma$ credible level. The constraints from VLBA+Gaia DR1 are less powerful because of the smaller number of sources.
Note that in this context, the energy density values can take negative values, simply due to our definition in Eq.~\eqref{eq:omega_u}. This arises because $V$ denotes the difference between left- and right-handed polarizations; specifically, $V = I$ corresponds to fully left-handed polarization, while $V = -I$ corresponds to fully right-handed polarization.

\begin{table}[htbp]
    \centering
    \begin{tabular}{|c|c|c|c|}
        \hline
        Dataset & $h_{70}^2\Omega_{U}^{2\sigma}$ & $h_{70}^2\Omega_{V}^{2\sigma}$ & $h_{70}^2\Omega_{\rm gw}^{\rm up, 95\%}$ from~\cite{Jaraba:2023djs} \\\hline
        Gaia DR3 & $-0.007\pm 0.035$ & $-0.020\pm 0.025$ & 0.087 \\
        VLBA & $0.000\pm 0.010$ & $-0.004\pm 0.010$ & 0.024 \\
        VLBA+Gaia DR1 & $-0.004\pm 0.015$ & $-0.007\pm 0.014$ & 0.034 \\\hline
    \end{tabular}
    \caption{The $2\sigma$ credible intervals for the distributions shown in Figs.~\ref{fig:intersection}-\ref{fig:VLBA_Gaia}. For reference, the 95\% credible upper bounds for the parity-even case $\Omega_{\rm gw}$ computed in~\cite{Jaraba:2023djs} are also provided in the last column.}
    \label{tab:results}
\end{table}

\section{Discussion and conclusion}
\label{sec:conlusion}
In this paper, we have used the Gaia DR3 and VLBA datasets to obtain the first constraints on parity-violating signals in the underlying SGWB. 
Although neither experiment provides time-series data at this stage, they can still be used to set an upper limit on the SGWB amplitude. The constraint is valid for frequencies between $H_0\sim10^{-18}\,{\rm Hz}$ and $1/T_{\rm obs}\sim 10^{-8}\,{\rm Hz}$ because the linear proper motion is induced by the quadrupole moment.
We have derived the relationship between the $EB$ correlation of the proper motion and a parity-violating amplitude, and have computed the corresponding probability density distributions by seeking this signal in the datasets. The tightest constraint arises from the VLBA data, which yields $h_{70}^2\Omega_V = -0.004\pm 0.010$ at the $2\sigma$ level. Since PTA observations are not sensitive to a parity-violating signal in an isotropic background, our results provide complementary information to PTA measurements.
This work should provide a template for such analyses, and we look forward to future future time-series data releases. Finally, while this work was being performed, a new approach that uses the Hellings-Downs curve of the data~\cite{Darling:2024myz} was proposed to further refine the parity-even astrometric constraints. Implementing such a method for a parity-violating SGWB could potentially improve our constraints by a similar factor (2-3), a direction we leave for future investigation.

\section*{Acknowledgments}

S.J. acknowledges support from the Agence Nationale de la Recherche (ANR) under contract ANR-22-CE31-0001-01. S.K. is supported by the I+D grant PID2023-149018NB-C42, the Consolidaci\'on Investigadora 2022 grant CNS2022-135211, and the Grant IFT Centro de Excelencia Severo Ochoa No CEX2020-001007-S funded by MCIN/AEI/10.13039/501100011033, the Leonardo Grant for Scientific Research and Cultural Creation 2024 from the BBVA Foundation, and Japan Society for the Promotion of Science (JSPS) KAKENHI Grant no. JP20H05853, and JP23H00110, JP24K00624. Q.L. is supported by World Premier International Research Center Initiative (WPI Initiative), MEXT, Japan. M.X.L. is supported by funds provided by the Center for Particle Cosmology at the University of Pennsylvania. The work of M.T. is supported in part by US Department of Energy (DOE) (HEP) Award DE-SC0013528.

This work has made use of data from the European Space Agency (ESA) mission {\it Gaia} (\url{https://www.cosmos.esa.int/gaia}), processed by the {\it Gaia} Data Processing and Analysis Consortium (DPAC, \url{https://www.cosmos.esa.int/web/gaia/dpac/consortium}). Funding for the DPAC has been provided by national institutions, in particular the institutions participating in the {\it Gaia} Multilateral Agreement.

\appendix
\section{Corner plot of MCMC runs}
\label{app:corner}

\begin{figure*}[htbp]
    \centering
    \includegraphics[width=\linewidth]{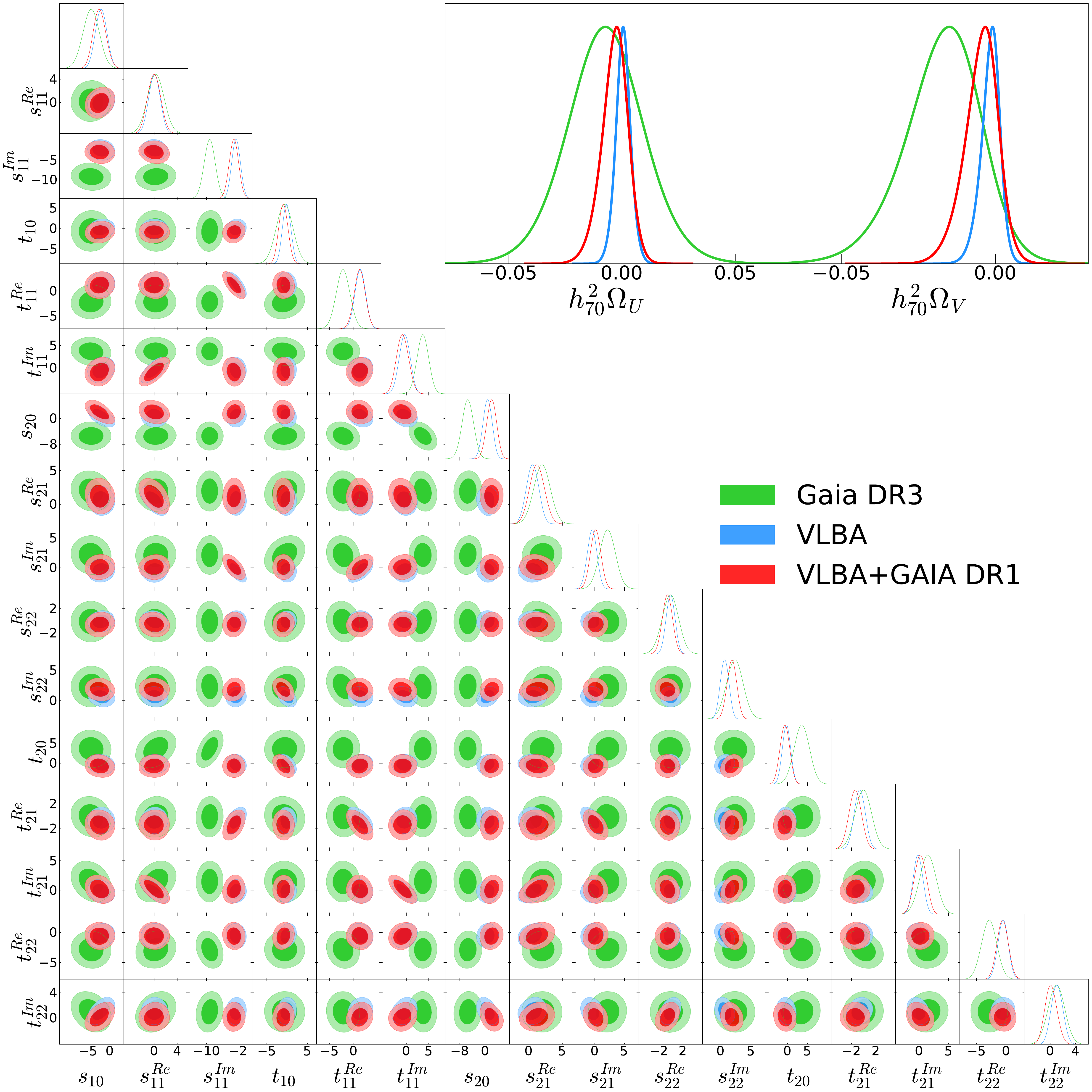}
    \caption{Corner plots for the three datasets considered in this article, together with the posterior probability distributions for $\Omega_{U}$ and $\Omega_{V}$.}
    \label{fig:corner}
\end{figure*}

In this Appendix, we provide the corner plots corresponding to the posterior distributions obtained in~\cite{Jaraba:2023djs} for the three datasets considered in this article. This information is summarized in Fig.~\ref{fig:corner}, together with the posterior probability distributions for $\Omega_{U}$ and $\Omega_{V}$. Here, we use the notation $s_{lm}\equiv\delta\dot{n}_{Elm}$, $t_{lm}\equiv\delta\dot{n}_{Blm}$, matching that in~\cite{Jaraba:2023djs} and~\cite{Darling:2018hmc}. 

\bibliography{Bibliography} 
\end{document}